# Microwave Photonic Multi-Mode Injection-Locked Frequency Divider With a Wide Operational Range Based on an Optoelectronic Oscillator

Siyu Liu, Kaitao Lin, Weiye Hu, Zhenzhao Yi, Xinhuan Feng, Jianghai Wo, and Jianping Yao, *Fellow, IEEE*

*Abstract*—We propose and implement a microwave photonic multi-mode injection-locked frequency divider (ILFD) with a wide frequency operational range based on an optoelectronic oscillator (OEO). In the OEO, a Mach-Zehnder modulator (MZM) and a photodetector (PD) are employed to construct a frequency multiplier to achieve an *N*-1 times frequency multiplication, which is then mixed with an external injection signal at an electrical mixer in the OEO loop. By adjusting the round-trip gain and time delay of the OEO loop, a radio frequency (RF) signal with a frequency that is 1/*N* that of the injection signal is generated, thus *N* times frequency division is achieved. Theoretical analysis and experimental verification are conducted to evaluate the effectiveness of the proposed ILFD. The results demonstrate that the system can divide a RF signal from 2.6 to 20.8 GHz to 1.3 to 1.95 GHz with different frequency division factors ranging from 2 to 13. A significant improvement in phase noise of 35.11 dB is also obtained at a frequency offset of 100 kHz when the frequency division factor is 13.

## I. INTRODUCTION

FREQUENCY dividers are essential components in microwave systems, such as radar and wireless communication systems for tasks such as clock recovery, signal processing, and frequency synthesis [1-4]. These frequency dividers can be classified as digital frequency dividers and analog frequency dividers. Digital frequency dividers are typically composed of triggers and logic gates and offer a flexibly tunable frequency division factor compared to analog frequency dividers. However, their operating frequency is often limited to several GHz, making them unsuitable for high-frequency systems [5]. On the other hand, analog frequency dividers can operate at much higher frequencies which are widely used in microwave systems. In general, analog frequency dividers can be further divided into regenerative frequency dividers and injection-locked frequency dividers (ILFDs). Compared to regenerative frequency dividers, ILFDs are considered the best pre-scalars due to their unique characteristics of high operating frequency (up to tens of GHz), low power consumption, and simple structure [6]. Generally, ILFDs can be single-mode ILFDs and multi-mode ILFDs with, respectively, a fixed and adjustable frequency division factor. However, a single-mode ILFD has a low frequency division factor (typically divide-by-2), which is primarily due to the inability to either generate ideal high-order harmonics or achieve second-mode harmonics with a sufficiently high intensity [7]. In recent years, there have been multiple reports on expanding the operational range and frequency division factor for single-mode frequency dividers. For example, the use of a dual-resonance resonator can enhance the operational range [8]. The use of a double cross-coupled CMOS LC-tank oscillator can also achieve a higher frequency division factor [9]. Other methods including a divide-by-4 ILFD based on harmonic boosting [10] and a divide-by-5 ILFD based on harmonic mixing [11] have also been demonstrated to have an increased frequency division factor. But all these methods are still limited to a single-mode frequency division with a maximum division factor of 5. On the other hand, a frequency synthesizer that can operate in multiple frequency bands has become an important design issue, which inevitably requires a multi-mode frequency divider with a frequency division factor that can be tunable to support a wide operational range [7]. Although there have been a few reports on multi-mode ILFDs in recent years, for example, a dual-mode ILFD that can support frequency generation at two bands of 38 and 57 GHz [6], a multi-mode ILFD incorporating a triple-band phase-locked loop (PLL) covering three bands of 40, 60, and 80 GHz [12], a divide-by-2/-3 dual-mode ILFD [7], and a divide-by-3/-5 dual-mode ILFD [13], most of them are dual-mode ILFDs with a narrow operational range, facing the same problem as single-mode ILFDs. A multi-mode frequency divider with a widely adjustable frequency division factor and a wide operational range can effectively alleviate the pressure and system complexity of later-stage digital frequency dividers, while expanding the operating frequency range of an entire PLL system.

This work was supported in part by Guangdong Province Key Field R&D Program Project under Grant 2020B0101110002, in part by National Key Research and Development Program of China (2021YFB2800804), and in part by the National Natural Science Foundation of China under Grants 61860206002. *(Siyu Liu and Kaitao Lin contributed equally to this work) (Corresponding author: Jianghai Wo)*

Siyu Liu, Kaitao Lin, Weiye Hu, Zhenzhao Yi, Xinhuan Feng, and Jianghai Wo are with the Guangdong Provincial Key Laboratory of Optical Fiber Sensing and Communications, Institute of Photonics Technology, Jinan University, Guangzhou 511443, China, and also with the College of Physics & Optoelectronic Engineering, Jinan University, Guangzhou 510632, China (e-mail: wojianghai@jnu.edu.cn).

Jianping Yao is with the Microwave Photonics Research Laboratory, School of Electrical Engineering and Computer Science, University of Ottawa, Ottawa, ON K1N 6N5, Canada (e-mail: jpyao@uottawa.ca).

In recent years, microwave frequency dividers based on photonic-assisted techniques have attracted widespread attention due to their advantages of high operating frequency, wide bandwidth, and immunity to electromagnetic interference [14, 15]. Various photonic-assisted frequency dividers have been demonstrated, including frequency dividers based on semiconductor lasers [16, 17], Fabry-Perot laser diodes (FP-LD) [18], semiconductor optical amplifiers (SOA) [19] and optoelectronic oscillators (OEOs) [20-28]. Among these frequency dividers, a microwave frequency divider based on an OEO has attracted much attention due to the advantageous features of high frequency, wide frequency tunability, and ultra-low phase noise. Although a regenerative frequency divider based on an OEO can achieve a multi-mode operation with a wide operational range, the frequency division factor is still limited to below 6. In addition, the phase noise of the frequency divided signal in a regenerative frequency divider only depends on that of the injection signal, which means the ultra-low phase noise characteristics of an OEO cannot be fully utilized to further reduce the phase noise of a frequency division signal at a far-carrier frequency offset range. A few ILFDs based on an OEO have been demonstrated in recent years, such as a single-mode divide-by-2 ILFD operated at 20 GHz [20], a multi-mode divide-by-2 to divide-by-6 ILFD with a narrow operational range based on an optical frequency comb incorporated in the OEO [22], a wide operational range but single-mode divide-by-2 or -3 ILFD based on a tunable microwave photonic filter [27].

In this paper, we propose and experimentally demonstrate a microwave photonic multi-mode ILFD with a wide operational range based on an OEO. In the OEO, a Mach-Zehnder modulator (MZM) and a photodetector (PD) are employed to construct an equivalent frequency multiplier to achieve an $N$-1 times frequency multiplication, which is then mixed with an external injection signal at an electrical mixer. The frequency division with a division factor of $N$ can be achieved when the phase condition is satisfied. The proposed approach is analyzed theoretically and demonstrated experimentally. In the experiment, a multi-mode ILFD that can operate from 2.6 to 20.8 GHz, with a tunable division factor from 2 to 13 is demonstrated. The phase noise of the frequency-divided signal is improved by 35.11 dB, at a far-carrier frequency offset of 100 kHz when the frequency division factor is 13. Moreover, the frequency tunable range and the tunable division factor for a frequency-fixed injection signal are also investigated.

## II. PRINCIPLE

Injection locking is essentially a nonlinear phenomenon in which oscillation signal track the frequency and phase of the injection signal [29]. For an oscillator, when a signal with a frequency close to the that of the free-running oscillation signal is injected, the oscillator is injection locked and the oscillation signal will approach the external injection signal. In an ILFD, the oscillation signal of an oscillator is frequency multiplied, and then mixed with an external injection signal. The frequency of the mixed signal is equal to the difference between the frequency of the external injection signal and the frequency-multiplied signal, which is used to lock the free-running oscillation signal, thereby realizing frequency division.

Fig. 1 illustrates the fundamental principle of the proposed multi-mode ILFD. Assume that a free-running oscillation signal at $f_{osc}$ is generated by the OEO, which is modulated on an optical carrier at an electro-optical modulator. After passing through a frequency multiplier to achieve $N$-1 times frequency multiplication, the optical signal is converted back to an electrical signal at a PD, which is then mixed with an external injection signal at an electrical mixer to produce an intermediate frequency (IF) signal with a frequency of $f_{inj}$-($N$-$1$)$f_{osc}$. If the frequency of the IF signal is close to the free-running oscillation signal $f_{osc}$, the free-running oscillation signal will be locked, and we have

$$f_{osc} = f_{inj} - (N-1)f_{osc} \qquad (1)$$

A frequency-divided signal is generated with its frequency given by

$$f_{out} = f_{osc} = \frac{f_{inj}}{N} \qquad (2)$$

The oscillation signal here is sent back to the electro-optic modulator to form a closed loop.

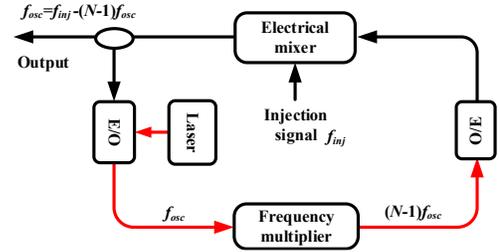

Fig. 1. The principle of the proposed multi-mode ILFD.

Fig. 2 illustrates the operational frequency range of a multi-mode ILFD. The blue region depicts the frequency range allowed for oscillation, and the green, red and yellow regions are the frequency ranges of its harmonics of different orders, which corresponds to the frequency tuning range of different operational modes. To achieve injection-locking-based frequency division under a given operational mode, the frequency of the injection signal needs to be within the corresponding frequency region. In this case, the frequency of an injection signal from different frequency bands can be divided into the same frequency range. Since no ultra-narrow band filter is employed, the adjacent harmonic regions may overlap, which allows for a wide and continuously adjustable operational frequency range of the proposed ILFD, as compared to that of other multi-mode ILFDs. Moreover, in the frequency overlapped region, the injection signal can be divided with different frequency division factors. Obviously, the larger the oscillation frequency range, the more the harmonic regions may overlap, which can realize a wider tuning range of the frequency division factor for a fixed injection frequency.

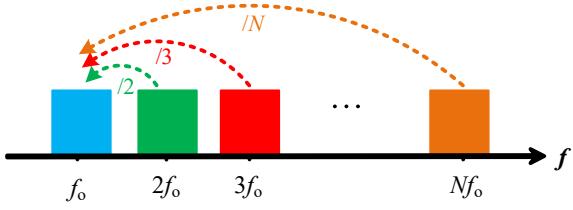

Fig. 2. Illustration of the operational frequency range of a multi-mode ILFD.

Fig. 3 shows the implementation of the proposed multi-mode ILFD system. The overall system is roughly the same as a regular OEO, with an additional electrical mixer inserted in the electrical path and an optical tunable delay line (OTDL) in the optical path. A laser diode (LD) is utilized to generate a continuous-wave (CW) light, which is then injected into a Mach-Zehnder modulator (MZM) through a polarization controller (PC), to which a microwave signal is applied via the RF port, to generate an optical signal with multiple sidebands of different orders. After being delayed by a single-mode fiber (SMF) and an OTDL, the optical signal is detected at a PD, and frequency-multiplied microwave signals are generated. Thus, the optical path is considered an equivalent frequency multiplier. An electronic amplifier (EA1) is connected after the PD to ensure that the frequency-multiplied signal before entering the electrical mixer has a sufficiently high power, which is then mixed with an external injection signal at the electrical mixer to generate an IF signal. A second electronic amplifier (EA2) is connected after the electrical mixer to compensate for the loss of the entire system. The signal at the output of EA2 is then split into two through an electrical coupler (EC), with one being fed back to the MZM to close the OEO loop, and the other being sent to an electrical spectrum analyzer (ESA) for spectrum and phase noise measurements.

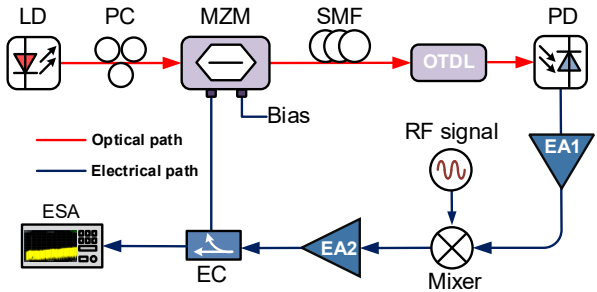

Fig. 3. The implementation of the proposed multi-mode ILFD. LD: laser diode, PC: polarization controller, MZM: Mach-Zehnder modulator, SMF: single-mode fiber, OTDL: optical tunable delay line, PD: photodetector, EA: electrical amplifier, EC: electrical coupler, ESA: electrical spectrum analyzer.

Assuming that the free-running oscillation signal in the OEO is $V_{osc}(t) = V_1 \cos(\omega_{osc} t + \theta_1)$, where $V_1$, $\omega_{osc}$, $\theta_1$ are the amplitude, frequency and phase of the oscillation signal, respectively. This signal is applied to the MZM to modulate the optical carrier. The optical signal at the output of the MZM can be expressed as

$$E_{out}(t) = \frac{1}{2} E_0 e^{j\omega_c t} \left[ e^{j\beta \cos(\omega_{osc} t + \theta_1)} + e^{-j\beta \cos(\omega_{osc} t + \theta_1) + j\phi} \right] \quad (3)$$

where $E_0$ and $\omega_C$ are the amplitude and frequency of the optical carrier, respectively, $\phi = \pi V_{DC} / V_\pi$ is a static phase due to the DC bias voltage applied to the MZM, $V_{DC}$ and $V_\pi$ are the bias voltage and half-wave voltage of the MZM, respectively, and $\beta = \pi V_1 / V_\pi$ is the modulation index. After passing through the SMF and the OTDL in the optical path, the optical signal is applied to the PD. The photocurrent at the output of the PD can be expressed as

$$i(t) = \frac{(1-\alpha)\eta E_0^2}{2} \{ 1 + \cos\phi \cos[2\beta \cos(\omega_{osc} t + \theta_1 - \omega_{osc}\tau)] \\ + \sin\phi \sin[2\beta \cos(\omega_{osc} t + \theta_1 - \omega_{osc}\tau)] \}$$

(4)

where $\alpha$, $\eta$ are the attenuation of the optical link and the responsivity of the PD, respectively, $\tau$ is the time delay of the optical path. According to the Jacobi Anger expansion, Eq. (4) can be further expressed as

$$i(t) = \frac{(1-\alpha)\eta E_0^2}{2} \{ 1 + J_0(2\beta) \cos\phi + \\ 2 \sum_{n=1}^{+\infty} J_n(2\beta) \cos(\phi - N\pi/2) \cos[N(\omega_{osc} t + \theta_1 - \omega_{osc}\tau)] \}$$

(5)

where $J_n(x)$ is the $N$-th order Bessel function of the first kind. After being amplified by EA1, the detected signal is mixed with the injection signal at the electrical mixer. Note that although there is no filter in the OEO loop, due to the frequency responses of the electronic and optoelectronic components, the system exhibits a low-pass frequency response. Therefore, we only retain the terms of the frequency difference between the injection signal and the ($N$-1)-th order frequency multiplied signal, while ignoring the sum frequency terms, the DC, and the mixing frequency terms between the injection signal and other harmonics at the output of the mixer. Thus, the IF signal fed back to the MZM can be expressed as

$$V(t) = G'_{N-1} \cos[\omega_{inj} t - (N-1)\omega_{osc} t + \theta_2 - (N-1)(\theta_1 - \omega_{osc}\tau)]$$

(6)

where $G'_N = [(1-\alpha)(1-\alpha')V_2 \eta R G E_0^2 J_n(2\beta) \cos(\phi - N\pi/2)]/2$, $\alpha'$ is the power loss introduced by the RF components, $G$ is the electrical gain of the EAs and $R$ is the resistance of the matched load, $V_2$, $\omega_{inj}$, $\theta_2$ are the amplitude, frequency and phase of the injection signal, respectively. According to the Barkhausen criterion [29], for a steady oscillation of the IF signal in the OEO, the following conditions should be satisfied

$$\begin{cases} \omega_{osc} = \omega_{inj} - (N-1)\omega_{osc} \\ \theta_1 = \left[\theta_2 + (N-1)\omega_{osc}\tau + 2k\pi\right]/N \ (k \in Z) \\ V_1 = G'_{N-1} \end{cases} \quad (7)$$

From Eq. (7), it can be seen that $\omega_{osc} = \dfrac{\omega_{inj}}{N}$, so a microwave signal with a frequency of $1/N$ of the injection signal will oscillate in the OEO by controlling the round-trip gain and the time delay of the loop. In addition, the phase noise of the oscillation signal is given by [30]

$$L(\Delta f) = \frac{L_{osc}(\Delta f)}{1 + (f_p/\Delta f)^2} + \frac{L_{inj}(\Delta f)/N^2}{1 + (\Delta f/f_p)^2} \quad (8)$$

where $L_{osc}(\Delta f)$ and $L_{inj}(\Delta f)$ are the phase noise terms of the free-running oscillation and the injection signal, respectively, $\Delta f$ is the frequency offset, and $f_p$ is a demarcation point in the offset frequency, which can be represented as [22]

$$f_p = \frac{\varepsilon^2 + \varepsilon\cos(\phi)}{(1 + \varepsilon\cos(\phi))^2} \frac{\omega_{osc}}{2Q} \quad (9)$$

where $\varepsilon = \sqrt{P_{IF}/P_{OSC}}$, $P_{IF}$ and $P_{OSC}$ are the powers of the IF and the free-running oscillation signal, respectively, $\phi = \arcsin(\Delta\omega_0/\Delta\omega_{lock})$ is their phase difference, $\Delta\omega_0$ is the angular frequency difference between the IF signal and the free-running oscillation signal. The locking range is $\Delta\omega_{lock} = (\omega_{osc}/2Q)\cdot(\varepsilon/\sqrt{1-\varepsilon^2})$, where $Q$ is the quality factor of the OEO loop. It can be inferred from Eq. (8) that within a near-carrier frequency offset range, the phase noise of the output signal is mainly determined by the injection signal. Theoretically, the phase noise performance will be improved by $20\log_{10}(N)$ dB for a frequency-divided microwave signal with a frequency division factor of $N$. However, at a far-carrier frequency offset range, the phase noise mainly depends on the free-running oscillation signal of the OEO.

### III. EXPERIMENT AND DISCUSSION

An experiment is performed based on the set up shown in Fig. 3 to evaluate the operation of the proposed multi-mode ILFD. A CW light at 1550.10 nm with a power of 7 dBm from the LD (Realphoton, TSL-E-C-96-N-FA) is launched into the MZM (Photoline, MX-LN-40), which has a 3 dB bandwidth of 40 GHz and a half voltage of 5.6 V. The SMF has a length of 430 m, which is relatively long to ensure that the OEO has a good phase noise performance. The OTDL (General Photonics, MDL-002) provides a continuously tunable optical time delay of up to 560 ps with a delay resolution of less than 1 fs is used to fine tune the loop delay, which allows easy adjustment of the phase to meet the conditions given in Eq. (7) in different frequency bands during the experiment, to achieve different frequency division factors. The PD (Realphoton, PIN/TIA20T) has a 3 dB bandwidth of 20 GHz and a responsivity of 0.8 A/W. EA1 (Realphoton, RFA-4X25-EVB) is used to provide an electrical gain of over 37 dB to ensure that the electrical signal has sufficiently high power for mixing. The electronic mixer (Magnum, MC57PG-2) is used to mix the frequency multiplied signal with the external injection signal from a microwave signal generator with a power of 0 dBm. EA2 (Realphoton, RFA-4X25-EVB) is used to compensate for the loss of the OEO loop. The amplified electrical signal is split into two by a 1-40 GHz bandwidth electrical coupler, with one path being fed back to the MZM, and the other sent to an ESA (Keysight, N9020B) for spectrum and phase noise measurement.

We first measure the spectrum of the free-running oscillation signal generated by the OEO without applying the injection signal. It is worth noting that although our system lacks a dedicated filter, the non-flat frequency response of the optoelectronic and electronic components within the entire loop forms a broadband low-pass filter. Therefore, all modes within this range have the potential to oscillate as long as the net gains of the modes exceed unity. The spectrum is shown in Fig. 4(a). The free-running oscillation signal at a frequency of 1.6 GHz is obtained. Fig. 4(b) is a zoom-in view of Fig. 4(a) with a frequency span of 40 MHz, it can be observed that there are multiple oscillation modes around 1.6 GHz. Due to the absence of a narrowband filter in the entire system, there is not a dominant mode in the cavity. The free spectral range (FSR) of the OEO is 460 kHz, indicating that the entire length of the OEO loop is approximately 434 m. Subsequently, we inject a microwave signal with a frequency of 3.2 GHz into the OEO and adjust the loop delay by tunning the OTDL to meet the divide-by-2 phase condition described in Eq. (7), to generate a divide-by-2 signal. As shown in Fig. 4(c), a stable divide-by-2 signal is generated. Fig. 4(d) is a zoom-in view of the generated signal, which indicates the OEO is operating in single mode. A further zoom-in view is shown in the inset of Fig. 4(d), by which the side mode suppression ratio (SMSR) is estimated, which is more than 64 dB. We further evaluate the stability of the frequency divider by measuring the power of the divide-by-2 signal for over 10 minutes. The power fluctuation is less than 0.03 dB, indicating the outstanding stability of the proposed ILFD.

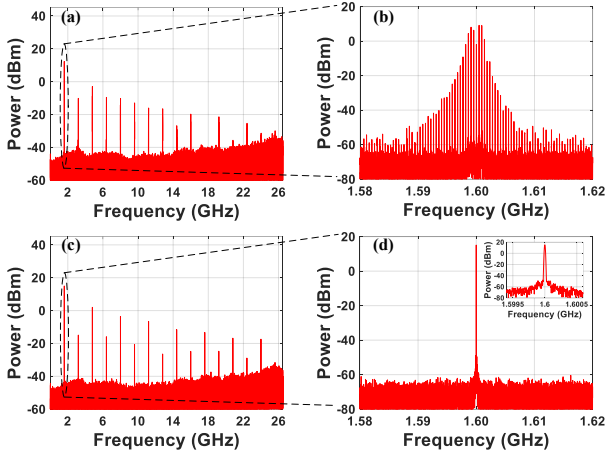

Fig. 4. (a) The spectrum of the free-running oscillation signal without an external injection signal. (b) Zoom-in view of the signal at 1.6 GHz with a span of 40 MHz and a RBW of 10 KHz. (c) The spectrum of the injection-locked frequency divided signal with an external injection signal of 3.2 GHz. (d) Zoom-in view of the frequency divided signal with a span of 40 MHz and a RBW of 10 kHz, the inset shows a further zoom-in view with a span of 1 MHz.

Fig. 5 shows the measured single-sideband phase noises of the free-running oscillation signal at 1.6 GHz, the external injection signal at 3.2 GHz, and the frequency divided signal at 1.6 GHz, which are measured by a built-in phase noise analyzer of the ESA. It can be clearly observed that within the frequency range near the carrier, the phase noise of the frequency divided signal is mainly determined by the external injection signal. However, when the frequency offset exceeds a demarcation point $f_p$ mentioned in Eq. (9), the phase noise is largely dependent on the free-running oscillation signal. The result in Fig. 5 demonstrates that the proposed ILFD can improve the phase noise of the injection signal. The phase noise is increased by 5.27 dB at 10 kHz frequency offset, as compared to that of the injection signal, which is close to the theoretical value of $20\log_{10}(N)$ dB for $N=2$. The phase noise is also improved thanks to the low phase noise feature of the OEO at a far-carrier frequency offset range. For example, the phase noise is 21.63 dB lower than that of the injection signal at the frequency offset of 100 kHz.

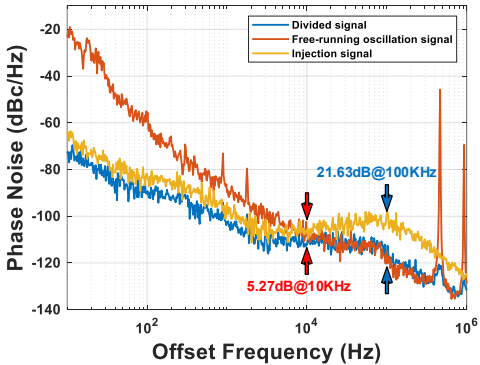

Fig. 5. The phase noise measurements of the free-running oscillation signal at 1.6 GHz, the injection signal at 3.2 GHz, and the frequency divided signal at 1.6 GHz.

For practical applications such as a pre-scalar, the signal under frequency division from different frequency bands are always divided to a certain frequency range for further digital process to relax the bandwidth requirement of the digital circuit, which means a multi-mode operation ability of the frequency divider. In traditional electronic multi-mode ILFDs, it is necessary to design harmonic enhancement modules or switch conversion circuits according to specific requirements, resulting in relatively complex circuits. However, in the proposed multi-mode ILFD, it is only necessary to adjust the loop delay through tuning the OTDL to achieve the frequency division of the injection signal. To demonstrate the multi-mode operation of the proposed multi-mode ILFD, different injection signals from 3.2 to 20.8 GHz are injected, with a step of 1.6 GHz. By adjusting the OTDL, all the injection signals from 3.2 to 20.8 GHz are divided to 1.6 GHz, corresponding to a frequency division factor from 2 to 13, as shown in Fig. 6.

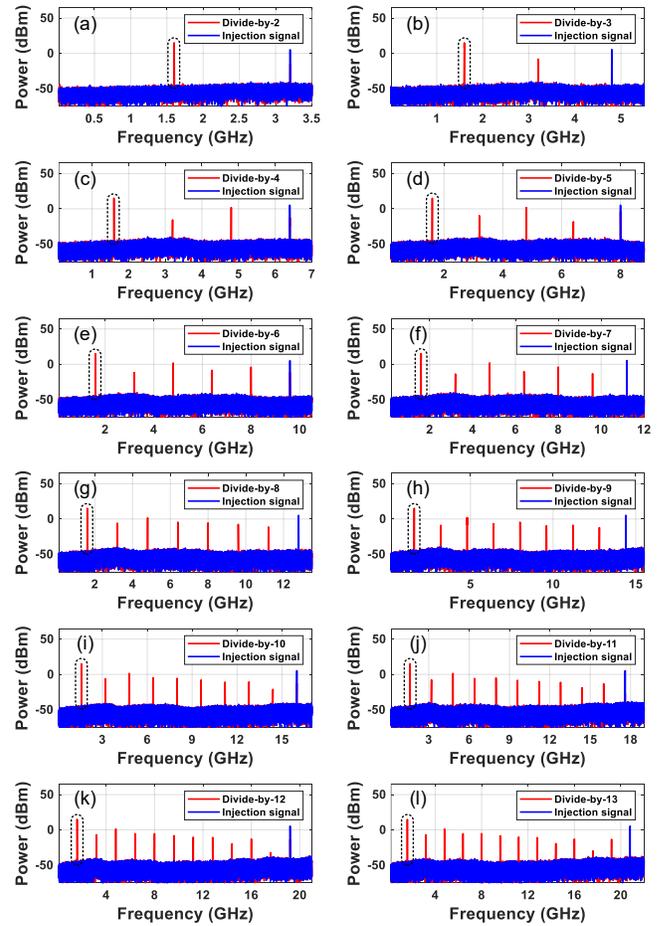

Fig. 6. The spectra of the injection and frequency divided signals when the injection signals are (a) 3.2 GHz, (b) 4.8 GHz, (c) 6.4 GHz, (d) 8.0 GHz, (e) 9.6 GHz, (f) 11.2 GHz, (g) 12.8 GHz, (h) 14.4 GHz, (i) 16.0 GHz, (j) 17.6 GHz, (k) 19.2 GHz, and (l) 20.8 GHz, corresponding to different frequency division factors from 2 to 13.

The corresponding phase noise results under different frequency division factors are shown in Fig. 7. The phase noise of the frequency divided signal at 1.6 GHz, the free-running oscillation signal at 1.6 GHz, and the injection signals at 3.2 GHz, 8 GHz, 14.4 GHz, and 20.8 GHz, are all presented, corresponding to the frequency division factors of 2, 5, 9, and 13, respectively. As can be seen, the phase noise of the injection signals is improved by 5.27 dB, 13.71 dB, 18.00 dB, and 21.24 dB at 10 kHz frequency offset, respectively. The improvements in the phase noise at 100 kHz frequency offset are 21.48 dB, 21.62 dB, 30.74 dB, and 35.11 dB, respectively. The experimental results of the phase noise improvement near the carrier frequency are basically consistent with the theoretical values, namely $20\log_{10}(N)$. At a far-carrier frequency offset range, the phase noise of the frequency divided signal tracks that of the OEO. The result demonstrates the superiority of the proposed multi-mode ILFD in reducing the phase noise of the injection signal. Note that the phase noise of an OEO is largely dependent on the optical delay of the loop, the phase noise of the frequency divided signal at far-carrier frequency offset will be further improved if a longer SMF is used.

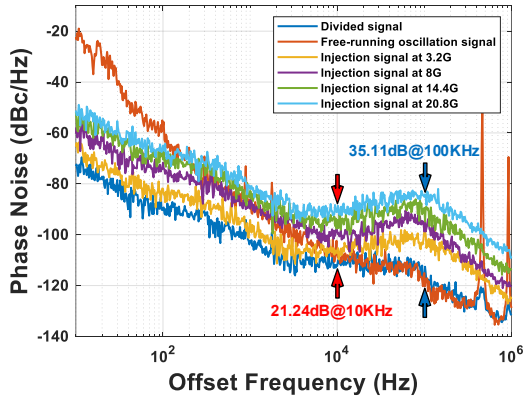

Fig. 7. The phase noise of the divide-by-$N$ signal at 1.6 GHz, the free-running signal at 1.6 GHz, and the injection signal at 3.2 GHz, 8 GHz, 14.4 GHz, and 20.8 GHz.

The locking range of the proposed multi-mode ILFD is also tested, which is done by slightly changing the frequency of the injection signal and adjusting the OTDL to meet the phase condition in Eq. (7). Fig. 8 shows the superimposed spectra of the divide-by-2 signals when the frequency of injection signal is tuned from 3.2 to 3.2009 GHz with a step of 100 kHz. The result demonstrates that injection-locked frequency division can be achieved across the entire FSR of the OEO.

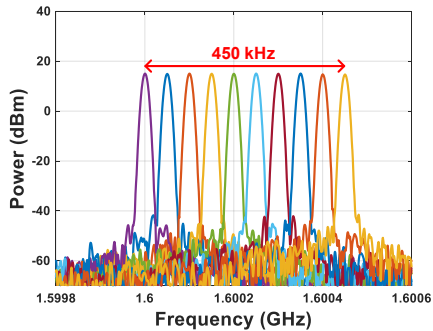

Fig. 8. The locking range test result. The RBW is 10 kHz.

The frequency tunning of the injection signal under a given frequency division factor is then investigated. Fig. 9(a) shows the superimposed spectra of the injection and divide-by-2 signals when the injection signals are sequentially injected from 2.6 to 3.8 GHz with a step of 600 MHz, while Fig. 9(b) shows the superimposed spectra of the injection and divide-by-3 signals when the injection signals are sequentially injected from 3.9 to 5.7 GHz with a step of 900 MHz. It is obvious that all the injection signals are frequency divided, and the power difference of the frequency divided signals is less than 1.6 dB. Note that this frequency tunning range can be further increased by utilizing optoelectronic and electronic components with flatter frequency responses. The results in Figs. 8 and 9 indicate that the proposed ILFD can achieve a continuously tunable frequency division of the injection signal over a wide frequency range.

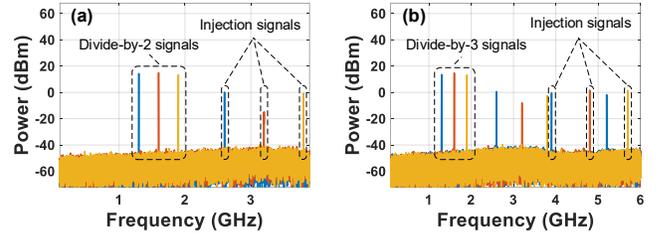

Fig. 9. (a) The superimposed spectra of the injection and divide-by-2 signals. (b) The superimposed spectra of the injection and divide-by-3 signals.

As demonstrated in Fig. 6, tunable frequency division factors are obtained by tuning the OTDL and changing the frequency of the injection signal. In fact, the frequency division factor under a fixed injection frequency can also be adjusted due to the frequency overlap of the frequency tuning range of different operational modes. When the injection signal falls within the frequency overlapped region, the time delay of the cavity can be adjusted to meet the phase condition for frequency division with different factors. Fig. 10 shows the superimposed spectra of the frequency divided signals with the frequency division factors of 2 and 3 when a 3.9 GHz signal is injected. As can be seen in Fig. 10, a divide-by-2 signal at 1.95 GHz or divide-by-3 signal at 1.3 GHz is obtained simply by adjusting the time delay of the loop.

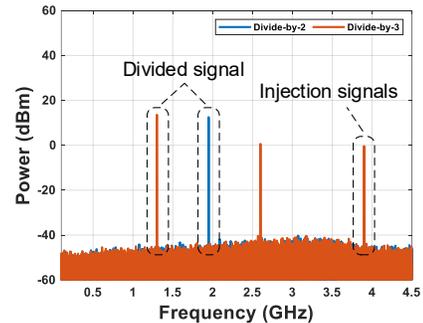

Fig. 10. The superimposed spectra of a divide-by-2 signal at 1.95 GHz and a divide-by-3 signal at 1.3 GHz with an injection signal of 3.9 GHz.

## IV. CONCLUSION

We proposed and implemented a microwave photonic multi-mode ILFD with a wide operational range based on an OEO. The key to this system is the frequency multiplication in the optical domain and the frequency mixing accomplished in the electrical domain. By adjusting the round-trip gain and

4delay of the loop, a microwave signal with a frequency that is 1/*N* that of the injection signal was generated. Theoretical analysis was conducted to obtain the frequency locking condition under different frequency division factors. Experimental results showed that the phase noise was reduced by up to 21.24 dB at 10 kHz frequency offset and 35.11 dB at 100 kHz frequency offset when the frequency division factor was 13. The results of the frequency locking range and the frequency tuning range demonstrated that the proposed ILFD could achieve a continuously tunable operation over a wide frequency range. The proposed multi-mode ILFD can effectively alleviate the pressure and system complexity of later-stage digital dividers, while expanding the operating frequency range of the entire PLL system.

## References

[1] Y. H. Chuang, S. H. Lee, S. L. Jang, J. J. Chao, and M. H. Juang, "A ring-oscillator-based wide locking range frequency divider," *IEEE Microw. Wireless Compon. Lett.*, vol. 16, no. 8, pp. 470-472, Aug. 2006.

[2] S. Pan and J. Yao, "Optical clock recovery using a polarization-modulator-based frequency doubling optoelectronic oscillator," *J. Lightwave Technol.*, vol. 27, no. 16, pp. 3531-3539, Aug. 2009.

[3] H. R. Erfani-Jazi and N. Ghaderi, "A divider-less, high speed and wide locking range phase locked loop," *AEU Int. J. Electron. Commun.*, vol. 69, no. 4, pp. 722-729, Apr. 2015.

[4] S. Lee, K. Takano, R. Dong, S. Amakawa, T. Yoshida, and M. Fujishima, "A 37-GHz-input divide-by-36 injection-locked frequency divider with 1.6-GHz lock range," in *Proc. Asian Conf. IEEE Solid-State Circuits*, Nov. 2018, pp. 219-222.

[5] S. L. Jang, W. C. Lai, G. Y. Lin, and C. Y. Huang, "Injection-locked frequency divider with a resistively distributed resonator for wide-locking-range performance," *IEEE Trans. Microw. Theory Techn.*, vol. 67, no. 2, pp. 505-517, Feb. 2019.

[6] H. K. Chen, H. J. Chen, D. C. Chang, Y. Z. Juang, Y. C. Yang, and S. S. Lu, "A mm-wave CMOS multimode frequency divider," in *Proc. ISSCC Dig. Tech Papers*, Feb. 2009, pp. 280-281.

[7] H. M. Cheema, X. P. Yu, R. Mahmoudi, P. T. Zeijl, and A. Roermund, "A dual-mode mm-wave injection-locked frequency divider with greater than 18% locking range in 65nm CMOS," in *IEEE MTT-S Int Microw. Symp. Dig.* May 2010, pp. 780-783.

[8] S. L. Jang, Y. J. Chen, C. H. Fang, and W. C. Lai, "Enhanced locking range technique for frequency divider using dual - resonance RLC resonator," *Electron. Lett*, vol. 51, no. 23, pp. 1888-1889, Nov. 2015.

[9] S. L. Jang, C.-F. Lee, and W.-H. Yeh, "A divide-by-3 injection locked frequency divider with single-ended input," *IEEE Microw. Wireless Compon. Lett.*, vol. 18, no. 2, pp. 142-144, Feb. 2008.

[10] L. Wu and H.C. Luong, "Analysis and design of a 0.6 V 2.2 mW 58.5-to-72.9 GHz divide-by-4 injection-locked frequency divider with harmonic boosting," *IEEE Trans. Circuits Syst. Regul. Pap.*, vol. 60, no. 8, pp. 2001-2008, Aug. 2013.

[11] S. L. Jang, G. Z. Li, and W. C. Lai, "Wide-locking range RLC-tank balanced-injection divide-by-5 injection-locked frequency dividers based on harmonic mixing," *IEEE Trans. Microw. Theory Techn.*, vol. 68, no. 3, pp. 894-903, Mar. 2020.

[12] H. K. Chen, T. Wang, and S. S. Lu, "A millimeter-wave CMOS triple-band phase-locked loop with a multimode LC-based ILFD," *IEEE Trans. Microw. Theory Techn.*, vol. 59, no. 5, pp. 1327-1338, May 2011.

[13] T. H. Huang, S. H. Li, P. K. Tsai, and C. C. Liu, "Reconfigurable CMOS divide-by-3/-5 injection-locked frequency divider for dual-mode 24/40 GHz PLL application," in *IEEE International Symposium on Radio-Frequency Integration Technology (RFIT)*, Nov. 2012, pp. 68-70.

[14] J. Yao, "Microwave photonics," *J. Lightw. Technol.*, vol. 27, no. 3, pp. 314-335, Feb. 2009.

[15] Y. Wang, X. Li, J. Wo, J. Zhang, A. Wang, and P. Du, "Photonic frequency division of broadband microwave signal based on a Fourier domain mode-locked optoelectronic oscillator," *Opt. Laser Technol.*, vol. 147, Art. no. 107704, Mar. 2022.

[16] S. C. Chan and J. M. Liu, "Microwave frequency division and multiplication using an optically injected semiconductor laser," *IEEE J. Quantum Electron.*, vol. 41, no. 9, pp. 1142-1147, Sept. 2005.

[17] L. Fan *et al.*, "Subharmonic microwave modulation stabilization of tunable photonic microwave generated by period-one nonlinear dynamics of an optically injected semiconductor laser," *J. Lightw. Technol.*, vol. 32, no. 23, pp. 4660-4666, Dec. 2014.

[18] M. Zhang, T. Liu, A. Wang, J. Zhang, and Y. Wang, "All-optical clock frequency divider using Fabry–Perot laser diode based on the dynamical period-one oscillation," *Opt. Commun.*, vol. 284, no. 5, pp. 1289-1294, Mar. 2011.

[19] A. Kelly, R. Manning, A. Poustie, and K. Blow, "All-Optical Clock Division at 10GHz and 20 GHz in a Semiconductor Optical Amplifier Based Nonlinear Loop Mirror," *Electron. Lett*, vol. 34, no. 13, pp. 1337-1339, Jan. 1998.

[20] H. Peng *et al.*, "Low phase noise 20 GHz microwave frequency divider based on a super-harmonic injection locked optoelectronic oscillator," in *IEEE International Frequency Control Symposium (IFCS)*, *IEEE*, May 2018, pp. 1-3.

[21] S. Liu, K. Lv, J. Fu, L. Wu, W. Pan, and S. Pan, "Wideband microwave frequency division based on an optoelectronic oscillator," *IEEE Photon. Technol. Lett.*, vol. 31, no. 5, pp. 389-392, Mar. 2019.

[22] Y. Xu *et al.*, "Injection-locked millimeter wave frequency divider utilizing optoelectronic oscillator based optical frequency comb," *IEEE Photon. J.*, vol. 11, no. 3, pp. 1-8, Jun. 2019.

[23] H. Zhou *et al.*, "Broadband two-thirds photonic microwave frequency divider," *Electron. Lett*, vol. 55, no. 21, pp. 1141-1143, Oct. 2019.

[24] Y. Chen, P. Zuo, T. Shi, and Y. Chen, "Photonic-based reconfigurable microwave frequency divider using two cascaded dual-parallel Mach-Zehnder modulators," *Opt. Exp.*, vol. 22, no. 21, pp. 30797-30809, Sept. 2020.

[25] S. Duan *et al.*, "Photonic-assisted regenerative microwave frequency divider with a tunable division factor," *J. Lightw. Technol.*, vol. 38, no. 19, pp. 5509-5516, Jun. 2020.

[26] H. Liu, N. Zhu, S. Liu, D. Zhu, and S. Pan, "One-third optical frequency divider for dual-wavelength optical signals based on an optoelectronic oscillator," *Electron. Lett*, vol. 56, no. 14, pp. 727-729, Jul. 2020.

[27] Y. Meng, T. Hao, W. Li, N. Zhu, and M. Li, "Microwave photonic injection locking frequency divider based on a tunable optoelectronic oscillator," *Opt. Exp.*, vol. 29, no. 2, pp. 684-691, Jan. 2021.

[28] H. Chen, P. Zhou, Z. Xu, C. Tao, H. Fu, Y. Zhou, and D. Chen, "Microwave Frequency Division Based on Optically Injected DFB-LD With an OEO Loop," *IEEE Photon. Technol. Lett.*, vol. 36, no. 3, pp. 147-150, Feb. 2024.

[29] S. Verma, H.R. Rategh, and T.H. Lee, "A unified model for injection-locked frequency dividers," *IEEE J. Solid-State Circuits*, vol. 38, no. 6, pp. 1015-1027, Jun. 2003.

[30] Z. Fan *et al.*, "Injection locking and pulling phenomena in an optoelectronic oscillator," *Opt. Exp.*, vol. 29, no. 3, pp. 4681-4699, Jan. 2021.
**Siyu Liu** received the B.E. degree in Optoelectronic Information Science and Engineering from Henan University of Science and Technology, Luoyang, China, in 2022. He is currently working toward the M.S. degree with the Institute of Photonics Technology, Jinan University, Guangzhou, China. His research focuses on microwave photonic signal generation and processing.

**Kaitao Lin** is currently pursuing the B.E. degree in optical engineering with the Institute of Photonics Technology, Jinan University, Guangzhou, China. His research interest is microwave photonic signal generation and processing.

**Weiye Hu** received the B.E. degree in Optoelectronic Information Science and Engineering from Jiangxi University of Science and Technology, Ganzhou, China, in 2022. He is currently working toward the M.S. degree with the Institute of Photonics Technology, Jinan University, Guangzhou, China. His research focuses on microwave photonic signal generation and processing.

**Zhenzhao Yi** received the B.E. degree in Optoelectronic Information Science and Engineering from Jiangsu University, Zhenjiang, China, in 2021. He is currently working toward the M.S. degree with the Institute of Photonics
7

Technology, Jinan University, Guangzhou, China. His research focuses on microwave photonic signal generation and processing.

**Xinhuan Feng** received the B.S. degrees in physics from Nankai University in 1995, and obtained the M.Sc and Ph.D degrees in 1998 and 2005, respectively from the Insititute of Modern Optics, Nankai University, China. From 2005 to 2008, she worked as a Postdoctoral Fellow at Photonics Research Centre of the Hong Kong Polytechnic University. Since March 2009, she has been with the Institute of Photonics Technology, Jinan University, China. Her research interests include various fiber active and passive devices and their applications, and microwave photonic signal processing.

**Jianghai Wo** received the B.S. and Ph.D. degrees in optical information science and technology, and optical engineering from Huazhong University of Science and Technology, Wuhan, China, in 2009 and 2014, respectively. In June 2021, he joined the Institute of Photonics Technology, Jinan University, Guangzhou, China, as an Associate Professor. His current research interests include microwave photonics and photonic radar system.

**Jianping Yao** (Fellow, IEEE) received the Ph.D. degree in electrical engineering from the Université de Toulon, Toulon, France, in December 1997. From 1998 to 2001, he was an Assistant Professor with the School of Electrical and Electronic Engineering, Nanyang Technological University (NTU), Singapore. In December 2001, he joined the School of Electrical Engineering and Computer Science, University of Ottawa, Ottawa, ON, Canada, as an Assistant Professor, where he was promoted to an Associate Professor in May 2003, and a Full Professor in May 2006. He was appointed as the University Research Chair of microwave photonics in 2007. In June 2016, he was conferred the title of Distinguished University Professor of the University of Ottawa. From July 2007 to June 2010 and from July 2013 to June 2016, he was the Director of the Ottawa-Carleton Institute for Electrical and Computer Engineering, Ottawa, ON, Canada. He is currently a Distinguished University Professor and the University Research Chair with the School of Electrical Engineering and Computer Science, University of Ottawa. He has authored or coauthored over 620 research papers including over 360 articles in peer-reviewed journals and over 260 papers in conference proceedings. Prof. Yao has served as a Committee Member for a number of international conferences, such as the IEEE Photonics Conference (IPC), Optical Fiber Communication Conference and Exposition (OFC), Bragg Gratings, Photosensitivity, and Poling in Glass Waveguides (BGPP), and International Topical Meeting on Microwave Photonics (MWP). He was a member of the European Research Council Consolidator Grant Panel in 2016, 2018, and 2020, the Qualitative Evaluation Panel in 2017, and a panelist of the National Science Foundation Career Awards Panel in 2016. He was an Elected Member of the Board of Governors of the IEEE Photonics Society from 2019 to 2021. He is also a Fellow of the Optical Society of America, the Canadian Academy of Engineering, and the Royal Society of Canada. He received the 2005 International Creative Research Award of the University of Ottawa. He was a recipient of the 2007 George S. Glinski Award for Excellence in Research. In 2008, he was awarded the Natural Sciences and Engineering Research Council of Canada Discovery Accelerator Supplements Award. He was selected to receive an Inaugural OSA Outstanding Reviewer Award in 2012. He was one of the top ten reviewers of IEEE/OSA JOURNAL OF LIGHTWAVE TECHNOLOGY from 2015 to 2016. He also received the 2017-2018 Award for Excellence in Research of the University of Ottawa. He was also a recipient of the 2018 R.A. Fessenden Silver Medal from IEEE Canada. He also serves as the Technical Committee Chair for IEEE MTT-S Microwave Photonics. He has also served as the Chair of a number of international conferences, symposia, and workshops, including the Vice Technical Program Committee (TPC) Chair of the 2007 IEEE Topical Meeting on Microwave Photonics, the TPC Co-Chair of the 2009 and 2010 Asia-Pacific Microwave Photonics Conference, the TPC Chair of the High-Speed and Broadband Wireless Technologies Subcommittee of the IEEE Radio Wireless Symposium 2009–2012, the Microwave Photonics Subcommittee of the IEEE Photonics Society Annual Meeting 2009, and the 2010 IEEE Topical Meeting on Microwave Photonics, the General Co-Chair of the 2011 IEEE Topical Meeting on Microwave Photonics, the TPC Co-Chair of the 2014 IEEE Topical Meetings on Microwave Photonics, the General Co-Chair of the 2015 and 2017 IEEE Topical Meeting on Microwave Photonics, and the General Chair of the 2019 IEEE Topical Meeting on Microwave Photonics. He was an IEEE MTT-S Distinguished Microwave Lecturer from 2013 to 2015. He is also a registered Professional Engineer of Ontario. He is also the Editor-in-Chief of IEEE PHOTONICS TECHNOLOGY LETTERS, a Former Topical Editor of Optics Letters, a Former Associate Editor of Science Bulletin, a Steering Committee Member of IEEE JOURNAL OF LIGHTWAVE TECHNOLOGY, and an Advisory Editorial Board Member of Optics Communications. He was a Guest Editor of a Focus Issue on Microwave Photonics in Optics Express in 2013, a Lead-Editor of a Feature Issue on Microwave Photonics in Photonics Research in 2014, and a Guest Editor of a Special Issue on Microwave Photonics in IEEE/OSA JOURNAL OF LIGHTWAVE TECHNOLOGY in 2018.